\documentclass[aps,12pt,final,notitlepage,oneside,onecolumn,nobibnotes,nofootinbib,superscriptaddress,noshowpacs,centertags]
{revtex4-2}

\usepackage{amsmath,amssymb}
\RequirePackage{mathtext}
\RequirePackage[utf8]{inputenc}
\usepackage{hyperref}
\usepackage{caption2}
\usepackage{graphicx}

\begin{document}

\title{Extensive Air Shower Registration\\at Two Depths with SPHERE-3 Detector}

\author{\firstname{V.~I.}~\surname{Galkin}}
\email{v\_i\_galkin@mail.ru}
\affiliation{Skobeltsyn Institute for Nuclear Physics, Lomonosov Moscow State University}
\affiliation{Faculty of Physics, Lomonosov Moscow State University}
\author{\firstname{E.~A.}~\surname{Bonvech}}
\affiliation{Skobeltsyn Institute for Nuclear Physics, Lomonosov Moscow State University}
\author{\firstname{D.~V.}~\surname{Chernov}}
\affiliation{Skobeltsyn Institute for Nuclear Physics, Lomonosov Moscow State University}
\author{\firstname{O.~V.}~\surname{Cherkesova}}
\affiliation{Skobeltsyn Institute for Nuclear Physics, Lomonosov Moscow State University}
\affiliation{Department of Space Research, Lomonosov Moscow State University}
\author{\firstname{E.~L.}~\surname{Entina}}
\affiliation{Skobeltsyn Institute for Nuclear Physics, Lomonosov Moscow State University}
\author{\firstname{V.~A.}~\surname{Ivanov}}
\affiliation{Skobeltsyn Institute for Nuclear Physics, Lomonosov Moscow State University}
\affiliation{Faculty of Physics, Lomonosov Moscow State University}
\author{\firstname{T.~A.}~\surname{Kolodkin}}
\affiliation{Skobeltsyn Institute for Nuclear Physics, Lomonosov Moscow State University}
\affiliation{Faculty of Physics, Lomonosov Moscow State University}
\author{\firstname{N.~O.}~\surname{Ovcharenko}}
\affiliation{Skobeltsyn Institute for Nuclear Physics, Lomonosov Moscow State University}
\affiliation{Faculty of Physics, Lomonosov Moscow State University}
\author{\firstname{D.~A.}~\surname{Podgrudkov}}
\affiliation{Skobeltsyn Institute for Nuclear Physics, Lomonosov Moscow State University}
\affiliation{Faculty of Physics, Lomonosov Moscow State University}
\author{\firstname{T.~M.}~\surname{Roganova}}
\affiliation{Skobeltsyn Institute for Nuclear Physics, Lomonosov Moscow State University}
\author{\firstname{M.~D.}~\surname{Ziva}}
\affiliation{Skobeltsyn Institute for Nuclear Physics, Lomonosov Moscow State University}
\affiliation{Faculty of Computational Mathematics and Cybernetics, Lomonosov Moscow State University}

\begin{abstract}
The progress in the development of the SPHERE-3 project is reported. The capabilities of the reflected Cherenkov light telescope and the direct light detector are stated. The procedures for separate EAS primary parameter assessment are mentioned. The advantage of dual atmospheric detection is underlined. An idea of the self consistent overall procedure is revealed.
\end{abstract}

\maketitle

{\let\thefootnote\relax\footnote{This a preprint of the Work accepted for publication in Physics of Atomic Nuclei, \copyright, copyright (2025), Pleiades Publishing, Ltd.}}\addtocounter{footnote}{-1}


\section{Introduction}
A new detector setup of the SPHERE series \cite{Chernov2022,Bonvech2023,Chernov2024} is currently undergoing the design and optimization stage. As stated in our previous papers, the main peculiarity of SPHERE-3 is going to be its comprehensive optimization towards the solution of the primary cosmic ray (PCR) mass composition problem. That mostly means that we admit a special difficulty of the problem which is usually underestimated. That also means we consider the vast computer simulations of the future detector behaviour to be the right way to find the best detector design and the most sensitive Cherenkov characteristics to measure, and to build a set of adequate data handling algorithms.

The original A.E. Chudakov’s idea~\cite{Chudakov1972} of an airborne detector of EAS Cherenkov light by itself makes our project peculiar:
\begin{itemize}
\item[-] the main telescope surveys a large patch (up to $\sim$1~km$^2$ in diameter) of snow surface, and the mosaic sees $\sim$50\% of this area;
\item[-] this makes it possible to estimate the primary energy and direction, the position of the shower axis on the snow, and the mass of the primary nucleus (roughly);
\item[-] changing the altitude of the installation allows you to adjust the sensitivity range for primary energies.;
\item[-] the setup is mobile and can be used in conjunction with ground-based installations that register EAS.
\end{itemize}

Still, these are not all the peculiarities available. SPHERE-1 and SPHERE-2 experiments used balloon-borne telescopes, SPHERE-3 will be carried by an unmanned aerial vehicle, which will make the upper hemisphere accessible to observations. It is well known that the Cherenkov light (CL) angular distribution bears important information on the shower longitudinal development. Our present studies show it is possible to use it for the enhancement of the primary mass recognition made by the reflected light telescope.

We hope that the future SPHERE-3 will be the first to realize the atmospheric registration of EAS at two depths, i.e. at the snow level and the altitude of the flying setup.

\section{EAS primary parameter definition by the SPHERE-3 detectors}

SPHERE-2 setup included only the reflected Cherenkov light (CL) telescope and was able to estimate primary energy, direction and location of the shower axis and primary particle mass. SPHERE-3 will also include such a telescope, though an optimized one. But it is destined for much more, i.e. to set a new standard in the primary mass assessment with the help of the detector of the direct CL. For this to come true we are going to focus on the events registered by both the lower (reflected light) and the upper (direct light) detectors. Paired EAS images considered together reveal different aspects of the event thus yielding more information on the shower than the images of the pair considered separately.

Let us now review the capabilities of the two detectors and the restrictions the physics of EAS puts on their dual detection.

\section{Reflected CL telescope}
A reflected light telescope (RLT) may be called the main detector because it provides the data for the estimates of the primary energy and the shower axis location and direction. Lateral distribution of light on its mosaic is also used for the primary mass assessment.

An EAS image on the RLT mosaic is a spatio-temporal distribution of photons (photoelectrons) which is integrated in time to form a lateral distribution. The latter is adjusted for the distortion of the RLT optics and then approximated by an axially symmetrical function. The resulting fit is used for the primary energy within 15--20\%~\cite{Chernov2024} accuracy and mass estimates (at the moment in the framework of a three-group classification scheme with characteristic misclassification errors of 0.30)~\cite{Chernov2024}. Maximum of the approximating function can be used as an estimate of a shower core location with 5 m error for 500 m flight altitude~\cite{Chernov2024}. Time delays of photons in different pixels of the mosaic are projected back to the snow to form an EAS light front representing the axis direction which is derived with an accuracy of 1.5--2.0$^\circ$~\cite{Chernov2024}.

Thus, RLT can yield a full set of the primary parameters of an event without any help from the direct light detector (DLD). Still, such help can be fruitful as we show below.

\section{Direct CL detector}

The DLD is going to register the angular distribution of light coming from EAS. It will be pointed at zenith, its field of view radius will amount to 20--25$^\circ$, in case of a single lens, or about 30$^\circ$, in case we use a hexagonal mosaic of 7 lenses working for one and the same sensor matrix. Currently we consider a configuration with a lens of area 400 cm$^2$.

\subsection{What showers can be clearly seen by the direct CL detector?}
With the field of view mentioned the detector will see the showers with zenith angle up to 20--25$^\circ$. Another restriction is ruled by the primary energy range and the detector area: the density of CL decreases with the core distance $R$, the amount of light collected by the sensor at $R>200$~m might be not enough for a comprehensive image processing, primarily for the mass assessment. At small $R$ the CL is abundant for the primary energy and detector area considered but the image dimensions also become small which increases the resolution requirements of the detector. As of now we assume the $R$ range for the direct CL detector measurements to be 100--200 m with its probable extension toward smaller $R$ in the future, though the region of $R<100$~m will not add too many events.

\subsection{What can direct light images add to the data given by the reflected light images?}

The DLD can hardly give an independent estimate of the primary energy nor the core location but it can substantially help in estimating the axis direction and the primary mass. The former possibility comes from the simple fact that the positions of the maximum and the center of mass of the CL angular distribution (and of the image in DLD) with respect to the shower axis depend on $R$ at the flight level. Knowing the estimates for the primary energy and the axis position at the flight level and direction by the data of RLT, it is possible to estimate the direction of the axis with smaller uncertainty of 0.4--0.5$^\circ$. That, in turn, makes it possible to improve the estimate of the axis position with respect to the detector at the flight level which is important for the construction of the system of criteria of the primary mass.

As mentioned above, the ability of the RLT to estimate the primary mass is limited and we have already learned the limit. Fortunately, here appears the ability of DLD. It is based on the remarkable properties of the spatio-angular distribution of EAS CL whose parameters depend on the primary mass. We have already studied the performance of the length of the angular distribution/image as a feature in different classification schemes and can share some conclusions.

\begin{figure}[t!]
\setcaptionmargin{5mm}
\onelinecaptionsfalse 
\includegraphics[width=1\linewidth]{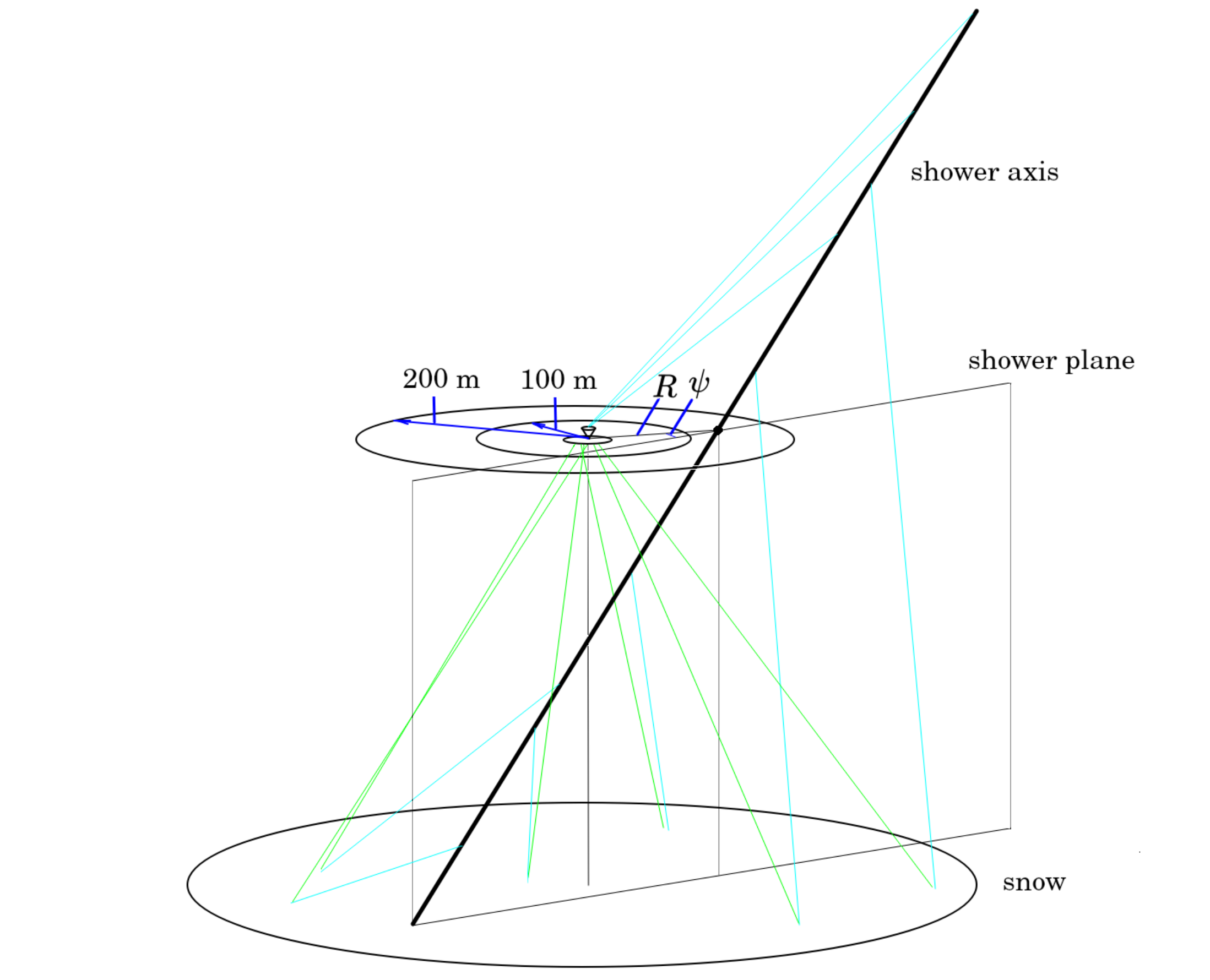}
\captionstyle{normal} \caption{Geometry of the detection at two levels. Cyan lines show direct CL photons, green ones depict reflected photons. The EAS axis must cross the ring at detector level and hit the snow within the circle for the dual detection to take place.}
\label{fig:geo}
\end{figure}

\begin{enumerate}
\item The length $a$ of the angular distribution/image (one of the Hillas’ parameters) is really sensitive to primary mass.
\item $a$ depends on the primary parameters (energy $E_0$, direction $\theta$, mass $A$) but also on the mutual geometry of the shower and the detector, i.e. core distance $R$ and azimuthal angle $\psi$ of the detector at the flight level (see~Fig.~\ref{fig:geo}). Naive classification scheme ignoring the mutual geometry doesn’t work.
\item A more sophisticated scheme incorporating a set of $a$-based criteria built each for its own $(R-\psi)$-bin really works, showing the misclassification errors lower than the scheme of RLT~\cite{Chernov2024}.
\item One can achieve even lower errors by applying absolute thresholds dependent on $R$ to the criteria of the scheme of para 3~\cite{Chernov2024}.
\end{enumerate}

Now we are studying another way to find a feature sensitive to the primary mass. An angular distribution/image can be sliced as long loaf, ratios of slice integrals are tried as candidates to the features for a classification scheme~\cite{Galkin2017}. More features can be found with accurate approximation of the angular distribution/image.

It seems clear now that the CL angular distribution/image can provide better separation of EAS by the mass of the primary particle. The most sensitive features F found will be used to construct a regression function A-F to ascribe the primary mass to individual showers.

Thus, the DLD supported by the RLT can enable us to move toward our main goal, i.e. the primary mass assessment. This is possible only for the showers registered by both detectors.
	 	 	 	
\subsection{What showers can be seen by both detectors?}
An EAS to be registered by the DLD and, after a short delay, by the RLT must satisfy the following conditions:
\begin{itemize}
\item[-] its axis must hit the snow within the area visible to RLT (snow core distance $R<175$~m for the setup flying at 500~m above the snow);
\item[-] the axis must lie within the ring 100~m~<~$R$~<~200~m at the flight level.
\item[-] its zenith angle should enable the EAS to be visible in DLD.
\end{itemize}

According to the results of our modelling, about 1/3 of the expected events will be detected by both RLT and DLD at the flight level 500 m. The dual detection fraction will decrease with height.

\subsection{How can the primary mass estimate benefit from the dual detection?}

\begin{figure}[t!]
\setcaptionmargin{5mm}
\onelinecaptionsfalse 
\includegraphics[width=\linewidth]{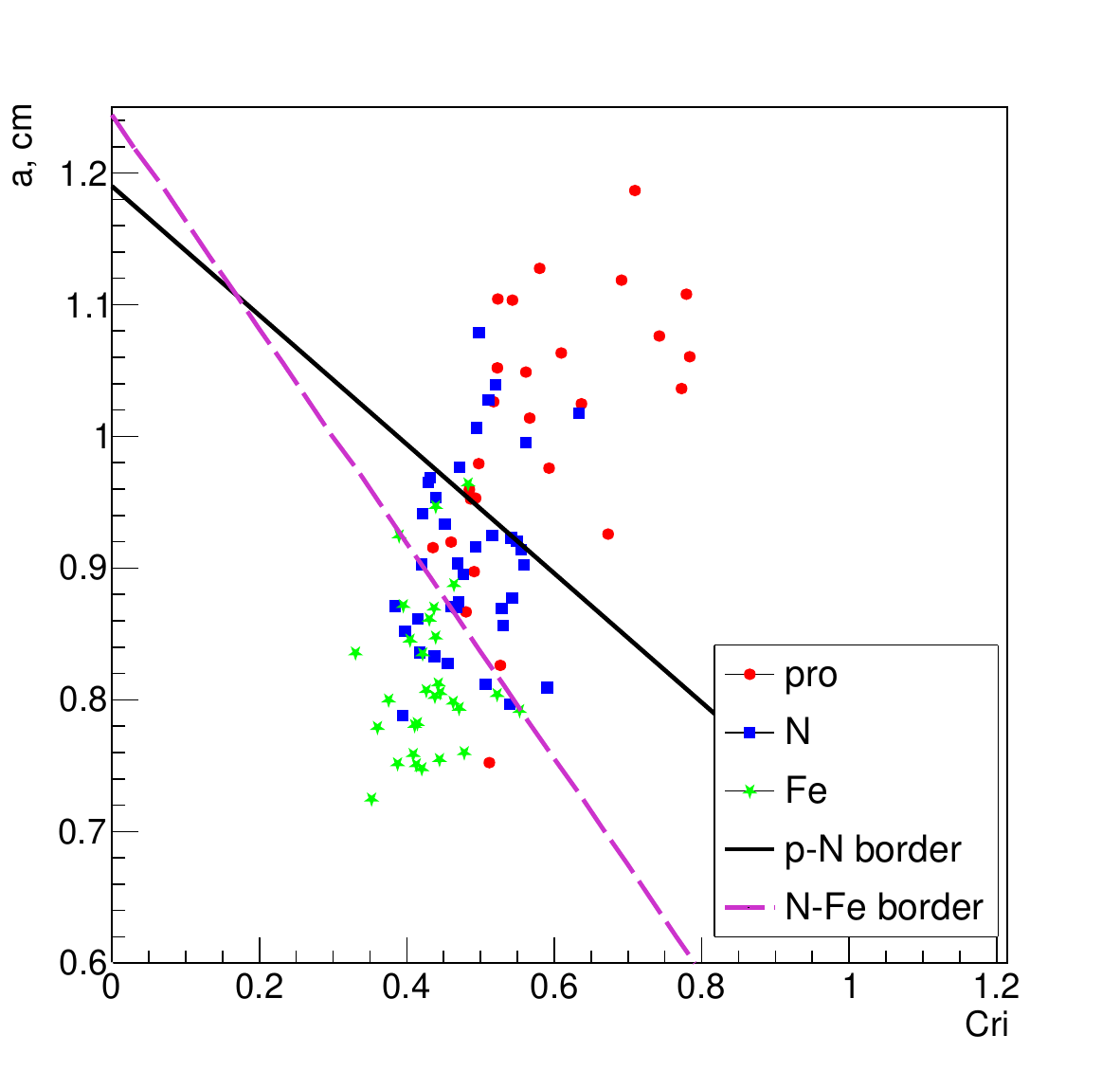}
\captionstyle{normal} \caption{2-dimensional classification scheme. Cri is the feature for RLT, $a$ is the length of the DLD image. The straight lines show the best borders between the pairs of classes p-N and N-Fe.}
\label{fig:dualcri}
\end{figure}

Encouraging results with the angular distribution/image shape features can be improved using a 2-dimensional scheme where one feature comes from RLT and the other from DLD. An example of such a criterion for one $(R-\psi)$-bin is shown in~Fig.~\ref{fig:dualcri}. Weak correlation of the features makes the separation substantially better: misclassification errors for both pairs p-N and N-Fe decrease to about 0.20~\cite{Galkin2025}.

\section{Outline of a procedure for self consistent primary parameter determination}
We have already seen the primary mass evaluation can benefit from the use of the data of both detectors. Similarly, the mutual geometry `shower---detector' can be known better if one uses in chain the procedures estimating the primary parameters by the RLT data and then those improving the values using the DLD data. The chain can be even turned in a loop.

\begin{figure}[t!]
\setcaptionmargin{5mm}
\onelinecaptionstrue 
\includegraphics{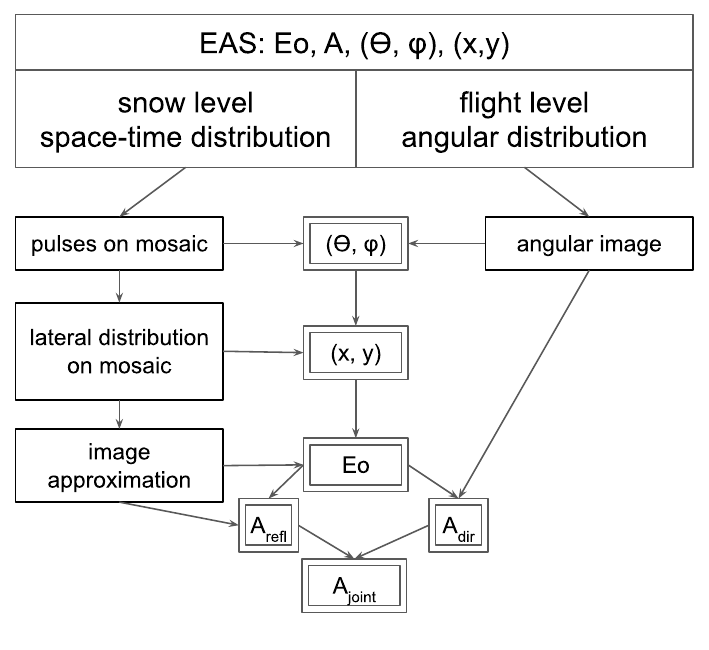}
\captionstyle{normal} \caption{Tentative scheme of a unified procedure for the primary parameter estimation.}
\label{fig:scheme}
\end{figure}

Summarizing this consideration, one can fuse all the specific primary parameter estimating procedures into a unified procedure impersonating a self consistent way of determination of the whole set of the primary parameters at once. A scheme of such unified procedure is given in~Fig.~\ref{fig:scheme}. Its left part presents the procedures dealing with the RLT data while the right one shows the ones handling the data of DLD. What the scheme lacks for the moment is a number of loops that will improve the primary parameter estimates reducing their uncertainties.	 	

The general procedure can be also thought of as a process of the multivariate fitting of a complex model, describing both images of the event, to the experimental data. But this model is still to be found. Anyway, such a procedure represents a target function which connects the input (measured quantities) to the output (the set of best estimates for the primary parameters).

\section{Regression versus classification}
We have already claimed that our goal is the best possible primary mass estimate and the way to achieve the goal is to construct a regression function A-F to mark each event with a mass tag. The classification schemes we presently use are just a tool to measure the sensitivity of the features to the mass. We already tried to build a regression scheme for the primary mass assessment for SPHERE-2 data~\cite{Latypova2023} and found rather big errors of the event-by-event mass estimates. Still, we stick to the idea of using a regression in mass at least for two reasons:

\begin{enumerate}
\item now we are going to use dual detection which will reduce the mass uncertainties;
\item even a mass estimate with some error is better than no mass estimate at all in the view of the self consistent procedure of the primary parameter evaluation.
\end{enumerate}

The latter fact will help to reduce the uncertainties of the whole set of primary parameters.

\section{Conclusion}
A new setup of the SPHERE series will incorporate a detector of CL angular distribution which will help the reflected light telescope in the determination of the EAS primary parameters.

With the currently accepted construction of the setup a substantial fraction of the detected events will yield decent images in both detectors.

A self consistent procedure for the determination of all primary parameters is needed to find the set of values with minimum uncertainties. As we care most of all about the primary mass estimates it is worthwhile to use a regression function for the mass evaluation inside the general procedure.

\begin{acknowledgments}
This work was supported by a grant from the Russian Science Foundation No 23-72-00006, https://rscf.ru/project/23-72-00006.
The research is carried out using the equipment of the shared research facilities of HPC computing resources at Lomonosov Moscow State University~\cite{SC}.
\end{acknowledgments}


\begin{thebibliography}{99}
\bibitem{Chernov2022}
D.V.~Chernov, C.G.~Azra, E.A.~Bonvech, V.I.~Galkin, V.A.~Ivanov, V.S.~Latypova, D.A.~Podgrudkov, T.M.~Roganova, Phys. At. Nucl. \textbf{85}(6), 641 (2022). https://doi.org/10.1134/S1063778822060059

\bibitem{Bonvech2023}
E.A.~Bonvech, C.J.~Azra, D.V.~Chernov, V.I.~Galkin, E.L.~Entina, V.A.~Ivanov, V.S.~Latypova, D.A.~Podgrudkov, T.M.~Roganova, M.D.~Ziva, Phys. At. Nucl. \textbf{86}(6), 1048 (2023). https://doi.org/10.1134/S1063778824010149

\bibitem{Chernov2024}
D.V.~Chernov, C.J.~Azra, E.A.~Bonvech, O.V.~Cherkesova, E.L.~Entina, V.I.~Galkin, V.A.~Ivanov, T.A.~Kolodkin, N.O.~Ovcharenko, D.A.~Podgrudkov, T.M.~Roganova, M.D.~Ziva, Phys. At. Nucl. \textbf{87}(S2), S319 (2024). https://doi.org/10.1134/S1063778824700959

\bibitem{Chudakov1972}
A.E. Chudakov, Proc. All-Union Symp. \textbf{620}, 69 (1972)

\bibitem{Galkin2017}
V.I.~Galkin, A.S.~Borisov, R.~Bakhromzod, V.V.~Batraev, S.~Latipova, A.~Muqumov, EPJ Web of Conf. \textbf{145}, 15004 (2017). https://doi.org/10.1051/epjconf/201714515004

\bibitem{Galkin2025}
T.A.~Dhzatdoev, E.A.~Bonvech, O.V.~Cherkesova, D.V.~Chernov, E.L.~Entina, V.I.~Galkin, V.A.~Ivanov, T.A.~Kolodkin, N.O.~Ovcharenko, D.A.~Podgrudkov, T.M.~Roganova and M.D.~Ziva, PoS (ICRC2025) \textbf{501}, 1357 (2025)

\bibitem{Latypova2023}
V.S. Latypova et al., Moscow State Univ. Bull. \textbf{78}(S1), S25 (2023). https://doi.org/10.3103/S0027134923070196

\bibitem{SC}
V.V.~Voevodin, A.S.~Antonov, D.A.~Nikitenko, P.A.~Shvets, S.I.~Sobolev, I.Yu.~Sidorov, K.S.~Stefanov, V.V.~Voevodin, S.A.~Zhumatiy, J. Supercomp. Frontiers and Innovations \textbf{6}(2), 4 (2019). https://doi.org/10.14529/jsfi190201

\end{thebibliography}
\end{document}